**Theory-training deep neural networks for an alloy solidification benchmark problem**

M. Torabi Rad[1], A. Viardin, G. J. Schmitz, and M. Apel

ACCESS e.V.

**Abstract**

Deep neural networks are machine learning tools that are transforming fields ranging from speech recognition to computational medicine. In this study, we extend their application to the field of alloy solidification modeling. To that end, and for the first time in the field, theory-trained deep neural networks (TTNs) for solidification are introduced. These networks are trained using the framework founded by Raissi et al. [1]–[3] and a theory that consists of a mathematical macroscale solidification model and the boundary and initial conditions of a well-known solidification benchmark problem. One of the main advantages of TTNs is that they do not need any prior knowledge of the solution of the governing equations or any external data for training. Using the built-in capabilities in TensorFlow, networks with different widths and depths are trained, and their predictions are examined in detail to verify that they satisfy both the model equations and the initial/boundary conditions of the benchmark problem. Issues that are critical in theory-training are identified, and guidelines that can be used in the future for successful and efficient training of similar networks are proposed. Through this study, theory-trained deep neural networks are shown to be a viable tool to simulate alloy solidification problems.

[1] Corresponding author: m.torabirad@access.rwth-aachen.de

1    Introduction

Machine learning is a branch of artificial intelligence that is based on the idea that computers can learn from data, identify patterns, and make predictions by minimal task-specific programming instructions. In the machine learning domain, neural networks are computing systems that consist of several simple but highly interconnected processing elements called neurons, which map an input array to one or multiple outputs. A schematic of a neural network is shown in Fig. 1 (a). In the figure, circles and arrows depict the neurons and connections between them, respectively. A neuron performs a weighted sum of the inputs $a_i$, adds the bias term $b$, and applies the non-linear function $g$ to the sum. The operations performed by a single node are summarized in Fig. 1(b) and can be mathematically expressed as

$$a_{out} = g(z) = g\left(b + \sum_{i=1}^{N} a_i w_i\right) \tag{1}$$

The neurons of a neural network are organized in layers. The first and last layers of a network and the layers between those two layers are termed the input, output, and hidden layers, respectively. A neural network that has three or more layers is termed a deep neural network. In the network sketched in Fig. 1, the first layer consists of two neurons only. For the hidden layers and the output layer, only the first two neurons and the last neuron are shown. The sketched network maps the input $X_i$, where $i$ represents a single data point (referred to as point hereafter), to $n$ different outputs $Y^j$ where $j = 1, \ldots, n$. The input and the outputs will be discussed in more detail in the context of theory-trained networks for solidification in the following sections. To make predictions using a neural network, the biases and weights of all the nodes first need to be calculated in a process called network training. Training starts by initializing those weights and biases to small and random values. The first estimates of the outputs are then calculated by forward passing (i.e., from the inputs to outputs) the information through the network. Those estimates are compared to the desired outputs, which can be the correct label of an image or an effectively zero error in satisfying an equation, to calculate a loss function. The loss is then minimized using an optimizer, which updates the weights by passing the information backward (i.e., from the outputs to the input layer). A forward pass followed by a backward pass is termed an iteration, and one iteration over the entire training dataset is called an epoch. After the training ends, the trained network can be used to make predictions by performing a single forward pass on new input data.

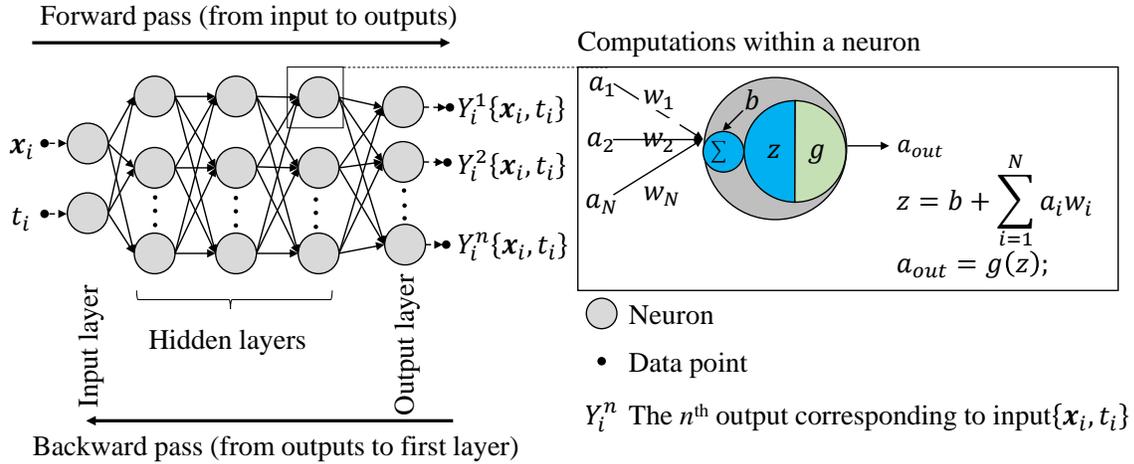

Figure 1. A schematic of a neural network that maps an input $X_i$ to outputs $Y^j$. Computations performed within a neuron are shown inside the box on the right. The network has three hidden layers and three nodes per hidden layer (i.e., $D = 3$ and $W = 3$)

Deep neural networks are now transforming fields such as speech recognition, computer vision, and computational medicine [4]. The main reason behind their remarkable achievements is the universal approximation theorem [5], [6]. According to that theorem, a neural network with even only one hidden layer can approximate any bounded continuous function as accurately as desired if it has enough nodes (i.e., is wide enough). In other words, neural networks are universal function approximators. However, representing a complex multi-variable function using a network with just one hidden layer is, in general, practically impossible. That difficulty can be overcome by adding more layers to the network (i.e., "going deeper"), which allows one to represent the same complex function with a drastically reduced number of total nodes [7]–[10].

The use of deep neural networks in the computational materials science domain is a very new area of research [11]–[14]. It is, however, interesting to note that shallow networks (networks with zero or one hidden layer) were applied in the field as early as the nineties [15], [16]. Deep neural networks have been so far used in the field to predict material properties (such as enthalpy of formation and Poisson's ratio) [17]–[19], effective macroscale properties of composites [20], and the strain field in a given microstructure [21], and to quantify [22], [23] and design [24], [25] microstructures. The relatively small number of studies in the literature is due to the novelty of the field, and more studies are expected to be performed in the near future.

Neural networks can also be trained to predict the solution of a Partial Differential Equation (PDE) or a system of coupled PDEs (the reason we do not use the term "solving a PDE" will be discussed at the end of section 2). The underlying reason for that capability is again the universal approximation theorem because the solution of a PDE is essentially a mapping between the independent variables (i.e., inputs) and the dependent variables (i.e., outputs). A neural network transforms the problem of numerically solving the equation into an optimization problem. Again, early work in the field goes back to the nineties, but those early works used only shallow networks. For example, Dissanayake and Phan-Thien [26] used a network with just one hidden layer to approximate the solutions of the linear Poisson equation and a heat diffusion equation. With the emergence of deep neural networks as an extremely powerful computational tool, and following the remarkable works in the research group of Karniadakis at Brown University, the topic has recently regained significant attention from different domains. Raissi et al. [1]–[3] set the foundations of a framework they termed physics-informed deep learning, details of which will be discussed in Section 2. The term "physics-informed" indicates that prior knowledge of the physics of a problem (for example, fluid flow around a cylinder), represented by equations (not to be confused with the solution of those equations), is used to train a network for that problem. As examples of highly relevant studies, Raissi and co-authors presented deep learning of vortex-induced vibrations [27], turbulent scalar mixing [28], and inverse problems involving PDEs [29]. Kharazmi et al. [30] developed a variational physics-informed neural network. Meng et al. [31] developed parallel physics-informed neural networks, and Chen et al. [32] developed Physics-informed neural networks for inverse problems. For more examples of relevant studies, the reader is referred to the review by Alber et al. [33].

The main objective of the present paper is to extend the application of deep neural networks and, more specifically, the framework which was founded by Raissi et al. [1]–[3] to the field of alloy solidification modeling by introducing TTNs for a well-known solidification benchmark problem. For pedagogical reasons, the theory-training process is discussed in detail. TTNs with different widths and depths are theory-trained, and their learning curves (i.e., how the value of the loss function changes during training) and predictions are analyzed. The results are used to identify challenges in theory-training networks for solidification and to give guidelines that can be used in future studies to select a proper value for the learning rate, which is the most important training hyper-parameter [4].

The paper is organized as follows: Section 2 details the essential ingredients of the theory-training framework. Sections 3 and 4 outline a macroscale model for solidification and the numerical solidification benchmark problem, respectively. Section 5 discusses how the model of Section 3 and the initial/boundary conditions of Section 4 are used to develop TTNs for the benchmark problem. Finally, in Section 6, the learning curves and predictions of TTNs trained using different learning rates are analyzed.

## 2 Theory-training neural networks

Training deep neural networks is a notoriously challenging task and typically requires large amounts of data, the lack of which is a factor that has so far limited the application of the networks in the computational materials science domain. Researchers have successfully trained networks using external sources of data (for example, X-ray images or results from Density Functional Theory simulations [12 and references therein]). Here, we use a training framework which was originally founded by Raissi et al. [1]–[3] and eliminates the need for any external source of data, be it experimental observations or results from other simulations, by relying on the ability of networks to learn the solution of partial differential equations. In this framework, a theory, which in the present paper is defined as the set of equations representing a theoretical (i.e., mathematical) model with a yet unknown solution and the initial/boundary conditions for a specific problem, is provided to a network. The network then essentially trains itself by learning the unknown solution of the model subject to the boundary and initial conditions of the problem. We, therefore, refer to these networks as theory-trained. Raissi and co-authors [27]–[29] have used the framework to successfully train networks for different interesting problems in fluid and quantum mechanics. They have referred to the framework using the term "physics-informed;" in the present study, however, we use the term "theory-trained" because, we believe, the latter term emphasizes that the core of the framework (i.e., training) is performed using theory, and not experimental measurements for example. Despite the remarkable achievements of Raissi and co-authors, numerous open questions still need to be addressed, especially if neural networks are to be used as a reliable alternative to traditional numerical methods to solve differential equations, such as finite difference or finite volume. For example, it is known that the learning rate is the most important hyper-parameter in training a network [4]. To the best of our knowledge, however, the effects of the learning rate on the training behavior and predictions of a theory-trained network has not been investigated in any detail. As another example, one can pose the following open question: is an increase in the width of a theory-trained network as beneficial as increasing its depth? This is a particularly interesting question because, in the literature, some studies suggest that increasing width is as or even more beneficial than increasing the depth [34][35][36], while others suggest that increasing depth is more beneficial [37]. Those studies, however, did not use theory-trained networks, and the question for that type of network has never been addressed before. Additional examples of open questions include: Is there a minimum depth/width below which a network will not be able to learn the physics incorporated in the theoretical model used for its training? And, what is the role of the size of the training dataset on the performance of a theory-trained network?

Further open questions arise if the theory-training framework is applied in the field of solidification. For example, to the best of our knowledge, in the literature, data-driven solutions of partial differential equations have been presented only for models where all the variables have an explicit relation to be calculated from. In the field of solidification, however, there exist commonly used mathematical models where some of the variables, solid fraction, for example, do not have explicit relations. The ability of the framework to train networks that can learn the solution of those types of models has not been tested yet. As another example, is deep learning of the high solid fraction stage of solidification as challenging as deep learning of the low solid fraction stage?

Let us elucidate the essential ingredients of theory-trained networks by first considering the example of a one-dimensional heat diffusion problem. The physical domain is bounded within $0 \leq x \leq x_{max}$, and we want to simulate this problem from $t = 0$ to $t = t_{max}$, where $t$ represents time. As shown in Figure 2, this problem is two-dimensional in the spatio-temporal domain. In the figure, the horizontal and vertical axes show the dimensionless time $t^* = t/t_{max}$ and distance from the origin of the physical space $x^* = x/x_{max}$, respectively.

Let us also recall that simulating this problem requires the heat diffusion equation, two boundary conditions (at $x^* = 0$ and 1) and an initial condition (at $t^* = 0$). Predicted temperatures are required to satisfy the diffusion equation and the boundary/initial conditions of the problem. For theory-trained networks, that requirement is met as follows. First, a set of points is distributed in the spatio-temporal domain. In Figure 2, a sample of these points is shown with the circles, upward/downward triangles, and squares, which are distributed randomly at $t^* = 0$ (the circles), $x^* = 0/1$ (the upward/downward triangles), and in the region that corresponds to $0 < x^* < 1; 0 < t^* < 1$ (the squares), respectively. We refer to these points as training points because they are relevant during training only. The circles, upward/downward triangles, and squares are used to train the network such that the temperatures that the network outputs during training (not to be confused with the temperatures that it predicts at the prediction stage) satisfy the initial conditions, boundary conditions at the bottom/top, and the diffusion equation, respectively. In other words, each training point is associated with a specific condition, and the trained network is required to satisfy all those conditions. This requirement is fulfilled in a procedure that is discussed next.

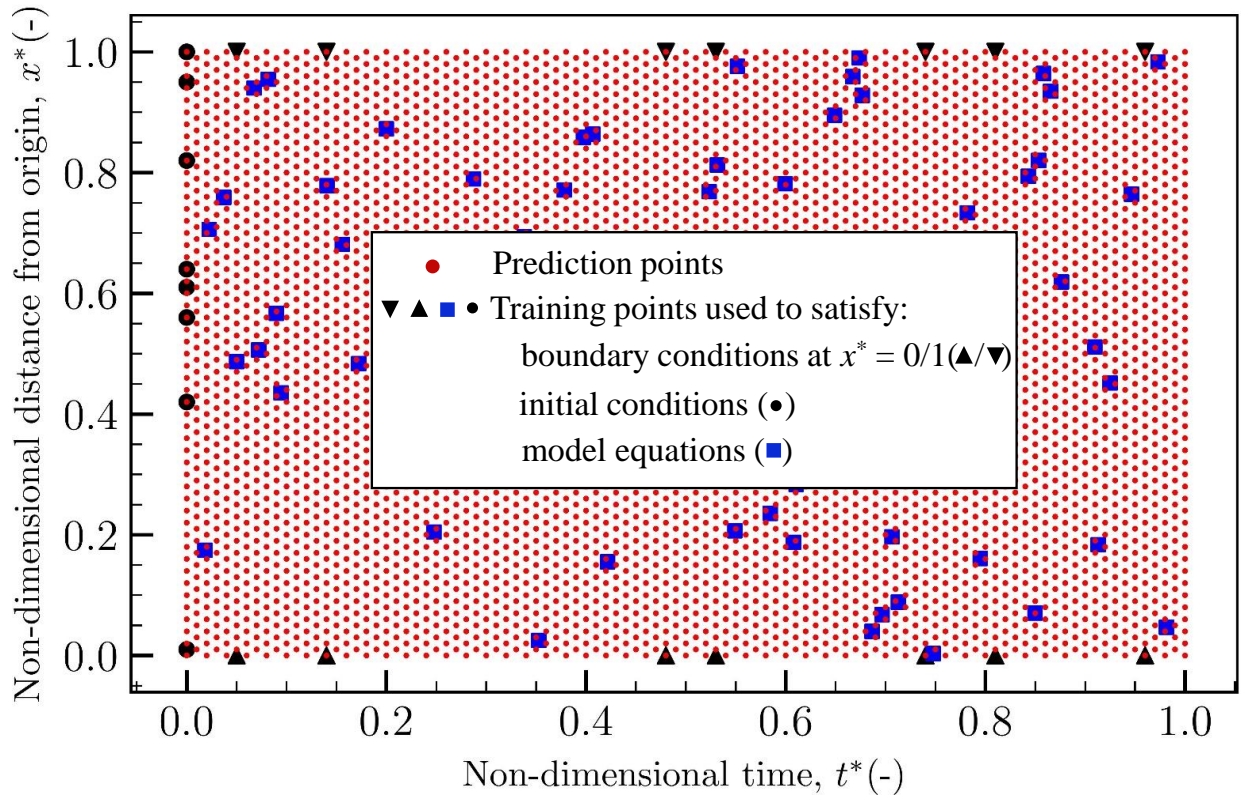

Figure 2. A schematic representation of a set of training points (upward/downward triangles, circles, and squares) and prediction points (red dots) for a problem that is two-dimensional in the spatio-temporal domain. The upward/downward triangles, circles, and squares are used to train the network such that its outputs satisfy the boundary conditions at $x^* =0/1$, initial conditions at $t^*= 0$, and also the governing equations of the theoretical model, respectively. The prediction points are comparable to nodes in a finite difference simulation.

During training, at each training epoch and for each training point, an error is calculated. This error is a measure of how well the condition corresponding to the training point is satisfied at the current epoch. For example, for the training points depicted in Figure 2 and our heat diffusion example problem, the error of a solid circle will be a measure of how well the temperature at the location of the circle satisfies the initial condition. Similarly, the error at the location of a triangle will measure how well the temperature at that location satisfies the boundary condition. For a square, the error is defined as the difference between the left-hand and right-hand sides of the heat diffusion equation and, therefore, will measure how well that equation is satisfied at the location of the square. The derivatives (temporal and spatial) that exist in the equation are calculated using a technique known as automatic differentiation. The details of the technique are beyond the scope of this paper and can be found elsewhere [38].

Once the errors for all the training points are calculated, the overall loss, which is defined as the mean of the squared errors, is calculated. The network is trained using an optimizer that changes the weights and biases of the nodes such that the value of the loss function decreases with training epochs. Training ends when the change in the loss between two consecutive epochs equals to zero (within numerical precision). If the final value of the loss is very small (for example, less than $10^{-5}$ if the inputs and outputs are all in the order of unity), the temperatures at every training point can be expected to satisfy the condition corresponding to that point.

The trained network can be used to make predictions. In the prediction stage, the network takes a set of new and yet-unseen points as input and, in our heat diffusion example problem, predicts the temperature for each point. In Fig. 2, these points are depicted by red dots. We refer to these points as prediction points because they are relevant during the prediction stage only. They are analogous to grid points in a finite difference method. There is, however, a fundamental and sharp difference between how predictions are performed with a theory-trained network and a finite difference method. In the latter, a prediction at any grid point requires a coding (i.e., human) instruction (for example, solving a discretized equation); in the former, however, no coding instructions are required in the prediction stage. Those instructions are required only during the training stage and at the training data points. In other words, the machine used the instructions provided to it during training to learn how to make predictions without any instructions. It should also be mentioned that because equations are not solved at the prediction points, we use the term "learning the solution of a PDE," instead of the term "solving a PDE." In fact, one of the main advantages of neural networks is that they can predict the solution of a PDE without solving it!

## 3    A model for macroscale solidification

Macroscale modeling of solidification typically relies on the concept of volume-averaging, which was first applied in the solidification field by Beckermann and Viskanta [39] in the mid to late 1980s and later significantly extended by Ni and Beckermann [40] and Wang and Beckermann [41]–[45]. In the present study, networks are trained using a volume-averaged model that disregards melt convection and assumes that solidification occurs in full equilibrium (i.e., lever solidification and zero nucleation undercooling), resulting in a single solid phase only (i.e., no eutectic phase). In this section, the equations of the model are outlined. Before proceeding, we emphasize that more realistic solidification models are available in the literature but are not used here because the present paper focuses on representing the first TTN for alloy solidification. Theory-training networks using more complex models represent a highly nontrivial task, which will be subject of a separate publication.

## 3.1 Dimensional form of the model

The model consists of equations for conservation of energy, average solute concentration in the solid, liquid, and mixture, and the liquidus line of the phase diagram. The energy equation reads [41]

$$\frac{\partial T}{\partial t} = \alpha_0 \nabla^2 T + \frac{h_{sl}}{c_p} \frac{\partial \phi_s}{\partial t} \qquad (2)$$

where $T$, $t$, $\alpha_0 = k/\rho c_p$, $h_{sl}$, $c_p$, and $\phi_s$ are the temperature, time, thermal diffusivity, latent heat, specific heat capacity, and solid fraction, respectively.

The equation for solute conservation in the liquid reads [41][46]

$$[k_0 \phi_s + (1 - \phi_s)] \frac{\partial C_l}{\partial t} = C_l (1 - k_0) \frac{\partial \phi_s}{\partial t} \qquad (3)$$

where $k_0$ and $C_l$ are the partition coefficient and solute concentration in the liquid, respectively. The above conservation equations need to be supplemented by the following thermodynamic relations

$$\begin{aligned} T > T_{liq}(C_l) &\rightarrow \phi_s = 0 \\ T_{sol}(C_l) < T \leq T_{liq}(C_l) &\rightarrow C_l = \frac{T - T_f}{m_l} \\ T \leq T_{sol}(C_l) &\rightarrow \phi_s = 1 \end{aligned} \qquad (4)$$

where $T_{liq} = T_f + m_l C_l$, $T_f$, $m_l$, $T_{sol} = T_f + m_l C_l/k_0$, and $k_0$ are the liquidus temperature of the alloy, melting temperature of the pure material, slope of the liquidus line of the phase diagram, solidus temperature, and partition coefficient, respectively. It should be mentioned that in a finite difference method enforcing different equalities in equation (4) is a trivial task (using "if" trees and few extra lines of code). With theory-trained networks, however, and as will be shown in the final section of the present paper, that task turns out to be highly non-trivial and requires careful selection of training hyper-parameters.

The equation for solute concentration in the solid reads

$$C_s = k_0 C_l \qquad (5)$$

Finally, the equation for the mixture (i.e., the region where solid and liquid coexist) solute concentration reads

$$C = \phi_s C_s + (1 - \phi_s) C_l \qquad (6)$$

There are two points about the model that require further attention and are discussed next. First, as discussed in Section 2, the previous studies used networks to learn the solution of a set of equations where all the unknown variables had an explicit relation to be calculated from. In the set of equations that we deep learn here (i.e., equations (2) to (6)), the solid fraction does not have an explicit relation; therefore, deep learning such a model is another novel aspect of the present study. Second, simulating this model using a finite difference method is now a trivial task in the literature;

however, as we will show in Section 6, deep learning of it turns out to be highly non-trivial. Finally, because the model disregards melt convection, it admits exact solutions for solid fraction and mixture concentration as: $g_s = (T - T_{liq})/[(1 - k_0)(T - T_f)]$ and $C = C_0$. These exact solutions will be used, at the prediction stage only, to evaluate the accuracy of the network predictions.

## 3.2 Scaled form of the model

In the model outlined in section 3.1, the values of different variables can differ from each other by orders of magnitude. For example, the temperature typically has values that are a few hundred degrees, while the values of the solid fraction are always between unity and zero. Having such a spread will typically lead to the failure of the training process (i.e., training will end with loss values that are not small enough). Therefore, before using the model to theory-train a network, it needs to be scaled such that all the variables have values with magnitudes ranging from zero to unity. To scale the model, the following scaled quantities are first defined

$$x^* = \frac{x}{x_{max}}; \quad t^* = \frac{t}{t_{max}}; \quad T^* = \frac{T - T_{liq}(C_0)}{T_{liq}(C_0) - T_{sol}(C_0)}; \quad C_l^* = \frac{k_0(C_l - C_0)}{(1 - k_0)C_0} \quad (7)$$

Substituting $x, t, T$, and $C_l$ appearing in equations (2) to (6) from equation (7) gives the following equations for the temperature, solute concentration in liquid, and the liquidus line of the phase-diagram

$$\frac{\partial T^*}{\partial t^*} = \frac{\alpha_0 t_{max}}{l^2} \nabla^{*2} T^* + \frac{h_{sl}}{c_p} \frac{1}{T_{liq}(C_0) - T_{sol}(C_0)} \frac{\partial \phi_s}{\partial t^*} \tag{8}$$

$$[k_0 \phi_s + (1 - \phi_s)] \frac{\partial C_l^*}{\partial t^*} = k_0 \left[1 + \frac{(1 - k_0)}{k_0} C_l^*\right] \frac{\partial \phi_s}{\partial t^*} \tag{9}$$

and

$$\begin{aligned} T^* &> 0 \rightarrow \phi_s = 0 \\ -1 &< T^* \leq 0 \rightarrow T^* + C_l^* = 0 \\ T^* &\leq -1 \rightarrow \phi_s = 1 \end{aligned} \tag{10}$$

4  A numerical benchmark problem for solidification

The solidification problem that is deep-learned in the present study is the numerical solidification benchmark problem introduced by Bellet el al. [47]. A schematic of the problem is shown in Figure 3. The problem consists of the solidification of a lead-18 wt. pct. tin alloy in a rectangular cavity. The cavity is insulated from the top and bottom. Initially, the melt is stationary, and its temperature is uniform and equal to the liquidus temperature at the initial concentration $C_0 = 18\ wt.pct$, 285.488 °C. Solidification starts by cooling the cavity from the sides through an external cooling fluid with the ambient temperature of $T_\infty = 25°C$ and overall heat transfer coefficient of $h_T = 400$ Wm$^{-2}$ K$^{-1}$. The width and height of the cavity are 0.1 and 0.06 cm, respectively. Due to the symmetry along the vertical mid-plane, only half of the cavity needs to be simulated. In the absence of melt convection, the problem is two-dimensional in the spatio-temporal domain. The thermophysical properties are outlined in [47] and are not repeated here for the sake of brevity. The initial and boundary conditions can be summarized as follows

$$T(x, t = 0) = T_{liq}(C_0); \quad C_l(x, t = 0) = C_0; \tag{11}$$

$$\nabla T(x = 0, t) = \frac{h_T}{k}(T - T_\infty); \quad \nabla T(x = w/2, t) = 0 \tag{12}$$

Expressing equations (11) and (12) in terms of the scaled variables introduced in equation (7) gives

$$T^*(x^*, t^* = 0) = 0; \quad C_l^*(x^*, t^* = 0) = 0 \tag{13}$$

$$\nabla^* T^*(x^* = 0, t) = Bi \left\{ \frac{T_{liq}(C_0) - T_\infty}{T_{liq}(C_0) - T_{sol}(C_0)} + T^* \right\}; \quad \nabla^* T^*(x^* = 1, t) = 0 \tag{14}$$

where $Bi = hw/(2k)$ is the Biot number. The scaled form of the model, outlined in equations (8) to (10), and the scaled form of the initial/boundary conditions of the benchmark problem, listed in equations (13) and (14), contain variables that have values between zero and unity and can, therefore, be used to theory-train a network for the benchmark problem.

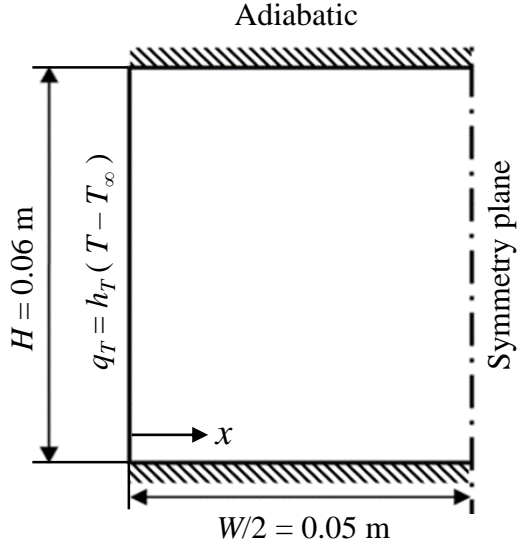

Figure 3. Schematic of the solidification numerical benchmark problem of Bellet et al.[46], [47]

5     <u>Theory-training networks for the solidification benchmark problem</u>

As already discussed in section 2, to theory-train a network, an error needs to be calculated for each equation that has to be satisfied and at every training point. These errors are defined as the difference between the left-hand and right-hand sides of the equations. For the solidification benchmark problem, these equations consist of equations (8) to (10) and equations (13) and (14); therefore, the errors are calculated from

$$E_T^i = \frac{\partial T^{*,i}}{\partial t^*} - \frac{\alpha_0 t_{max}}{l^2} \nabla^{*2} T^{*,i} + \frac{h_{sl}}{c_p} \frac{1}{T_{liq}(C_0) - T_{sol}(C_0)} \frac{\partial g_s^i}{\partial t^{*,i}} \tag{15}$$

$$E_{C_l}^i = [k_0 g_s^i + (1 - g_s^i)] \frac{\partial C_l^{*,i}}{\partial t^{*,i}} - k_0 \left[1 + \frac{(1-k_0)}{k_0} C_l^{*,i}\right] \frac{\partial g_s^i}{\partial t^{*,i}} \tag{16}$$

$$E_{liq}^i = \begin{cases} g_s^i & T^* > 0 \\ T^{*,i} + C_l^{*,i} & -1 < T^* \leq 0 \\ g_s^i - 1 & T^* \leq -1 \end{cases} \tag{17}$$

where the superscript $i$ denotes a training point that is located inside the spatio-temporal domain (i.e., not on the edges). Similarly, the errors in satisfying equations (13) and (14) are calculated from

$$E_{ic,T}^j = T^*(x^{*,j}, t^* = 0)$$
$$E_{ic,C_l}^j = C_l^*(x^{*,j}, t^* = 0) \tag{18}$$

where the superscript $j$ denotes a training point that is located on the axis $t^* = 0$. The errors in satisfying the boundary conditions are calculated from

$$E_{bc,1}^k = \nabla^* T^*(x^* = 0, t^{*,k}) - Bi \left\{ \frac{T_{liq}(C_0) - T_\infty}{T_{liq}(C_0) - T_{sol}(C_0)} + T^*(x^* = 0, t^{*,k}) \right\}$$
$$E_{bc,2}^k = \nabla^* T^*(x^* = 1, t^{*,k}) \tag{19}$$

where the superscript $k$ denotes a pair of training points that are located on axes $x^* = 0/1$ and have the same $t^*$. Finally, the loss is calculated from

$$Loss = \frac{1}{N_1} \sum_{i=1}^{N_1} \left[ (E_T^i)^2 + (E_{C_l}^i)^2 + (E_{liq}^i)^2 \right] + \frac{1}{N_2} \sum_{j=1}^{N_2} \left[ (E_{ic,T}^j)^2 + (E_{ic,C_l}^j)^2 \right]$$
$$+ \frac{1}{N_3} \sum_{k=1}^{N_3} \left[ (E_{bc,1}^k)^2 + (E_{bc,2}^k)^2 \right] \tag{20}$$

where again, $N_1$, $N_2$, and $N_3$ are the number of training points inside the spatio-temporal domain, on the axis $t^* = 0$, and axes $x^* = 0/1$, respectively; their sum (i.e., $N_1 + N_2 + N_3$) can be viewed as the size of the training data set. Note that all the terms on the right-hand side of equation (*20*) are calculated from equations with both sides of the order of unity, and this is a critical point in training a TTN.

It is now instructive to make a connection between this equation and the schematic shown in Figure 2. Inside the first summation on the right-hand side of the equation, as already discussed in connection with Equations (*15*)-(*17*), $E_T^i$, $E_{C_l}^i$, and $E_{liq}^i$ represent the errors in satisfying the equations for $T$, $C_l$, and liquidus line, respectively; they are evaluated at the location of the $i^{th}$ square in the figure. The first term on the right-hand side is the mean square of those errors evaluated at the locations of all the squares and measures how well the network satisfies the equations of the model at the training points. Inside the second summation, and as already discussed in connection with Equation (18), $E_{ic,T}^j$ and $E_{ic,C_l}^j$ represent the errors in satisfying the initial conditions for $T$ and $C_l$, respectively; they are evaluated at the location of the $j^{th}$ black circle in the figure. The second term is the mean square of those errors evaluated at the locations of all the black circles and measures how well the network satisfies the initial conditions. Similarly, inside the third summation, $E_{bc,1}^k$ and $E_{bc,2}^k$ are the errors in satisfying the bottom and top boundary conditions, evaluated at the location of the $k^{th}$ upward and downward triangles, respectively; the third term measures how well the network satisfies the boundary conditions at the training points.

During training, the weights of the nodes are updated using an optimizer such that the loss,

calculated from Equation (20), decreases with training epochs. The success of a training process can be easily evaluated by the final value of the loss. A value that is close to zero (between $10^{-5}$ and $10^{-6}$, for example) indicates that all the terms on the right-hand side of equation (20) are virtually zero and, therefore, the differential operator of the model and the initial/boundary conditions of the problem are accurately satisfied.

## 6   Results and Discussion

All the networks analyzed in the present study have hyperbolic tangent activation functions and were trained using the Adaptive Moment (Adam) optimizer [48] available in TensorFlow. For all the hyper-parameters, except the initial learning rate $\lambda_0$ of the optimizer, the TensorFlow default values were used; the value of $\lambda_0$ is however critical. It controls the stability and speed of the learning process. In the present study, networks were trained using different values of $\lambda_0$ and the effects of $\lambda_0$ on the training curves and predictions were analyzed to give guidelines that can be used to identify a suitable value in future studies.

Before proceeding, a minor difference between our TensorFlow implementation and the one exercised in the outstanding studies of Raissi et al. [1]–[3] should be mentioned. Their implementation used two networks, one to learn the hidden solution of an equation and the second one to enforce the structure of the equation; in our implementation, both of those tasks are performed using just one network.

### 6.1   Predictions of the reference TTN

The theory-training process and our implementation in TensorFlow are verified by analyzing the predictions of one of the TTNs as a reference. This reference network has nine hidden layers and four nodes per hidden layer (i.e., $D = 9$ and $W = 4$). A schematic of the reference network is shown in Figure 4(a). It was trained using a dataset that consisted of two-hundred points inside the spatio-temporal domain and twenty points on the $t^* = 0$ and $x^* = 0/1$ axes each (i.e., $N_1 = 200$, $N_2 = 20$, and $N_3 = 20$) all distributed randomly. Predictions were made into a 100×100 spatio-temporal grid.

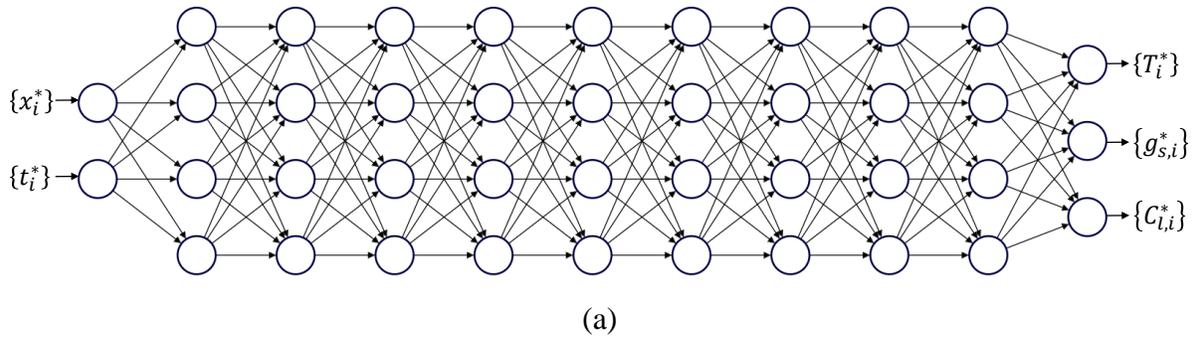

(a)

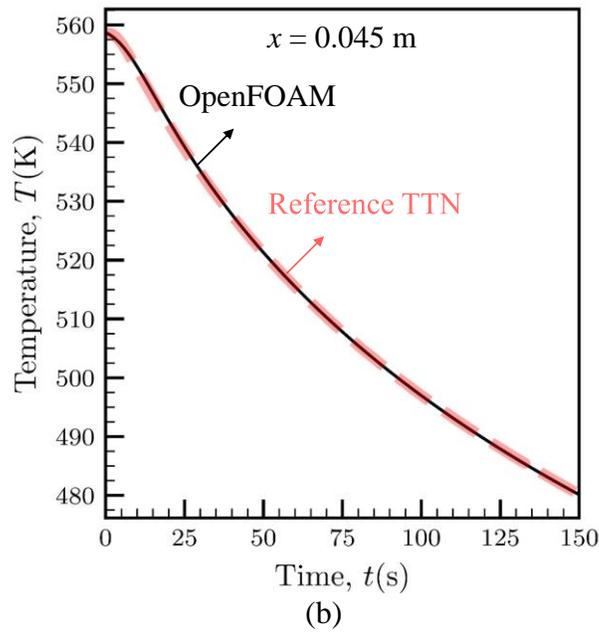

(b)

Figure 4. A schematic showing the architecture of the reference TTN, which has nine hidden layers and four nodes per hidden layer (a) and a comparison between temperatures predicted by that network (dashed red) and a previously developed OpenFOAM code [46], [49] (solid black). OpenFOAM results were used only to verify the predictions of the network and not for training. The displayed results are for location $x = 0.045$ m.

Predictions of the reference TTN are analyzed in two steps. In the first step, they are compared with the results obtained from numerically solving the equations of the model (i.e., equations (2) to (6)) using the traditional finite-volume method. The latter results were obtained using an OpenFOAM[50] code previously developed by Torabi Rad and Beckermann [46], [49]. It is emphasized that OpenFOAM simulations were performed only to verify the predictions of the reference TTN, and the results of those simulations were not used by any means in the theory-training process. That process, as already mentioned in Section 2, does not need any external sources of data. In the second step, contour plots of different variables (i.e., the temperature, solid fraction, liquid solute concentration and mixture concentration) predicted by the reference network are analyzed in detail to ensure that they satisfy the equations of the model and the initial/boundary conditions of the benchmark problem.

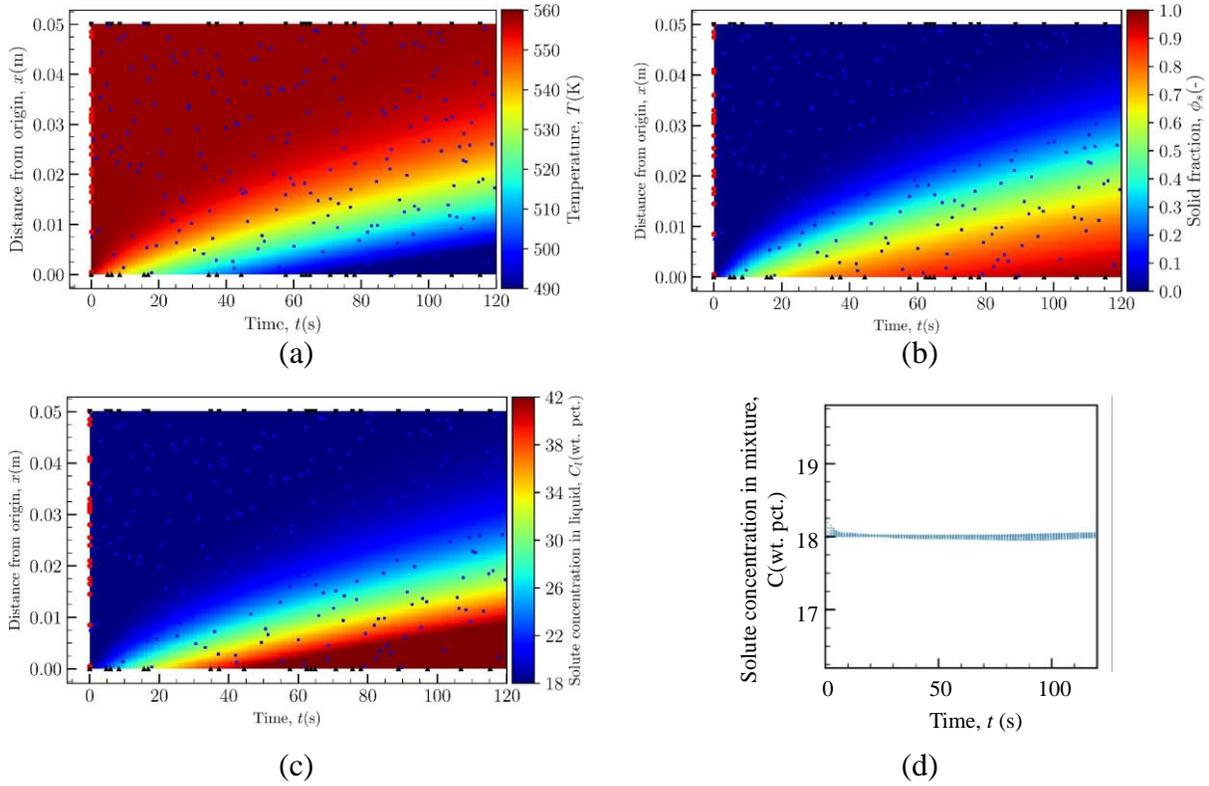

Figure 5. Contour plots showing the predictions of the reference TTN for temperature $T$ (a), solid fraction $\phi_s$ (b), and liquid concentration $C_l$ (c). The scatter plot in (d) shows mixture concentrations calculated from equations (5) and (6) using the predicted values for $C_l$ and $\phi_s$. The markers in plots (a) to (c) represent the training points.

In Figure 4(b), a comparison between the temperatures predicted by the reference TTN (dashed red) and those obtained from OpenFOAM simulations (solid black) is displayed. We iterate again that OpenFOAM results were not used for training the networks; they were used only to verify the predictions of the reference network. The figure shows the predicted temperatures at location $x = 0.045$ cm (i.e., $x^* = 0.9$) as a function of time, and the two predictions are virtually identical. Similar comparisons were made for the other variables predicted by the network, and those comparisons also resulted in excellent agreements. They are, however, not shown here for the sake of brevity. The observed agreement between the predictions of the reference network and OpenFOAM results was the first verification step. The second verification step is discussed next.

Figure 5 shows the values of the different variables predicted by the reference TTN. The color in Figure 5(a) to Figure 5(c) represents the temperature $T$, solid fraction $\phi_s$, and solute concertation in the liquid $C_l$, respectively; Figure 5(d) shows a scatter plot of the solute concentration in the mixture $C$ and each dot in the plot represents a single prediction point (out of 10000 points). In Figure 5(a) to Figure 5(c), the markers represent the training points. Next, these plots are examined in detail to ensure that the predictions of the reference network satisfy the equations of the model and also the initial and boundary conditions of the benchmark problem. From Figure 5(a) to Figure 5(c), it is evident that the values of $T$, $\phi_s$, and $C_l$ are initially, as expected, 558.638 K, 0, and 18, respectively. This reveals that the solution learned by the network satisfies the initial conditions. More importantly, recall from the discussion below equation (6) that, because the model disregards melt convection, the value of $C$ at every location in the space is expected to be equal to $C_0 = 18$. The scatter plot in Figure 5(d) shows that the reference network accurately satisfies this requirement, which further verifies the theory-training process and our implementation in TensorFlow. In the next Sub-section, the learning curves and predictions of networks with different widths and depths are analyzed in more detail.

## 6.2 Optimum initial learning rates for the Adam optimizer

Predictions of a neural network depend critically on how it is trained. Networks having the same architecture (i.e., the same depth $D$ and width $W$) but trained using different optimizers and/or different training hyper-parameters may yield entirely different predictions. As already discussed in Sub-section 6.1, in the present study, training experiments are performed using the Adam optimizer, which is one of the most commonly used optimizers in the deep learning community (as is indicated by more than seven thousand citations per year to the original paper[48]). The most important hyper-parameter of Adam optimizer is the initial learning rate $\lambda_0$[4]. The effects of the value of $\lambda_0$ on the training behavior and predicted results are discussed next in connection with Figure 6 and Figure 7.

Figure 6 shows the changes in the loss function during the training of networks all having $D = 12$ and $W = 4$, but trained using four different values of $\lambda_0$. The solid fractions and temperatures predicted for each $\lambda_0$ are shown in Figure 7. From Figure 6 it can be seen that with $\lambda_0 = 1 \times 10^{-5}$ and $3 \times 10^{-5}$ (the blue and black curves, respectively), the loss reaches a plateau (depicted by the dashed purple line in the figure) close to the end of training (i.e., epoch greater than 220000). With $\lambda_0 = 2.5 \times 10^{-6}$ or $\lambda_0 = 1.0 \times 10^{-4}$ (i.e., the red and green curves, respectively), however, this plateau is never reached. With $\lambda_0 = 2.5 \times 10^{-6}$, the loss keeps decreasing (but extremely slowly), even at very high epochs. With $\lambda_0 = 1.0 \times 10^{-4}$, the oscillations in the loss at high epochs become very significant (Note that there were some oscillations in the red, blue and black curves as well, but those oscillations are so insignificant that are not noticeable given the scale chosen for the vertical axis). In other words, the plateau in the loss is reached only for a range of $\lambda_0$ values; for a $\lambda_0$ below that range, the loss will keep decreasing extremely slowly even at high epochs and for a $\lambda_0$ above that range the loss will oscillate strongly.

By comparing the predicted $g_s$ and $T$ with the exact relationship between those two quantities displayed in Figure 7, it can be seen that only the networks that were trained with $\lambda_0 = 1.0 \times 10^{-5}$ or $\lambda_0 = 3.0 \times 10^{-5}$ are able to provide accurate predictions (see Figure 7(b) and (c)). The maximum/mean errors for these two networks are very low: 0.26/1.28 and 0.07/0.92 percent, respectively (see the contour plots in the lower-left corners, which show the percentage error in the predicted $C$, defined as $(C - C_0)/C_0$). Interestingly, in Figure 6, the loss plateaued for these two networks. For the network trained with $\lambda_0 = 2.5 \times 10^{-6}$ the predictions are accurate only at the high solid fraction region, i.e., $g_s > 0.6$ (see Figure 7(a)), and for the network trained with $\lambda_0 = 1.0 \times 10^{-4}$ predictions close to the liquidus temperature are entirely inaccurate as some of the solid fractions are negative! (see Figure 7(d) and the close up at the top-right corner). From the analysis of Figure 6 and Figure 7, it can be concluded that in order to have predictions that are accurate (i.e., satisfy the governing equations and the physical limits of the model variables), a network should be trained such that its training loss reaches a plateau close to the end of the training, and that can be achieved only when the value of $\lambda_0$ is within some intermediate range. We term that range the range of optimal learning rates and, in the next sub-section, provide a curve-fitting relation for its lower and upper limits.

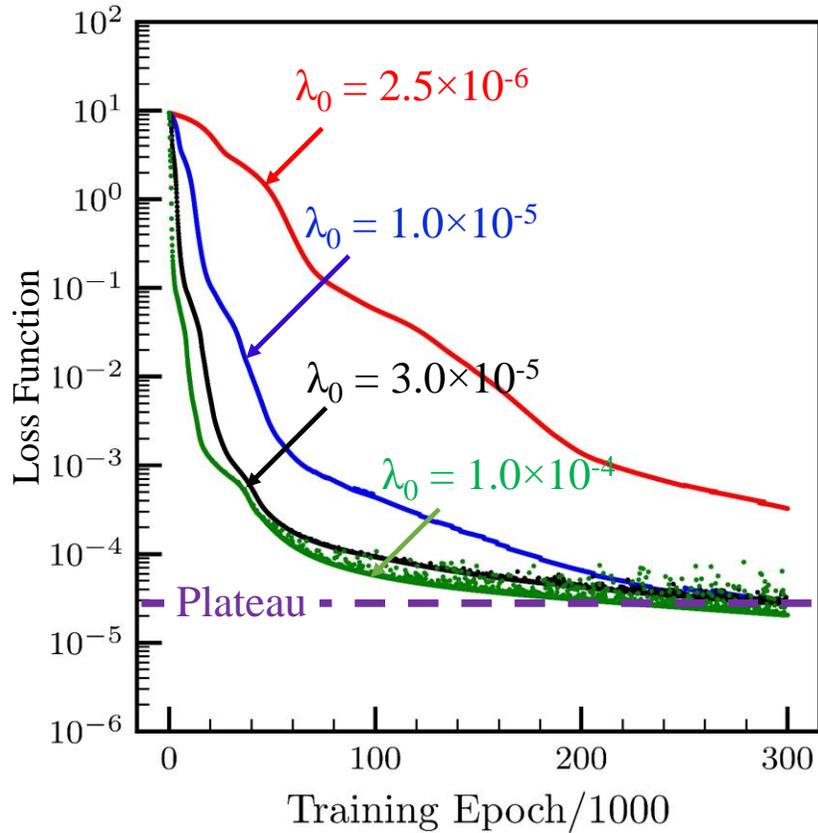

Figure 6. The changes in the value of the loss function during training displayed for four different networks that all have D = 12 and W = 4 but are trained using four different values of the initial learning rate $\lambda_0$. The solid fractions and temperatures predicted by these networks are shown in Figure 7.

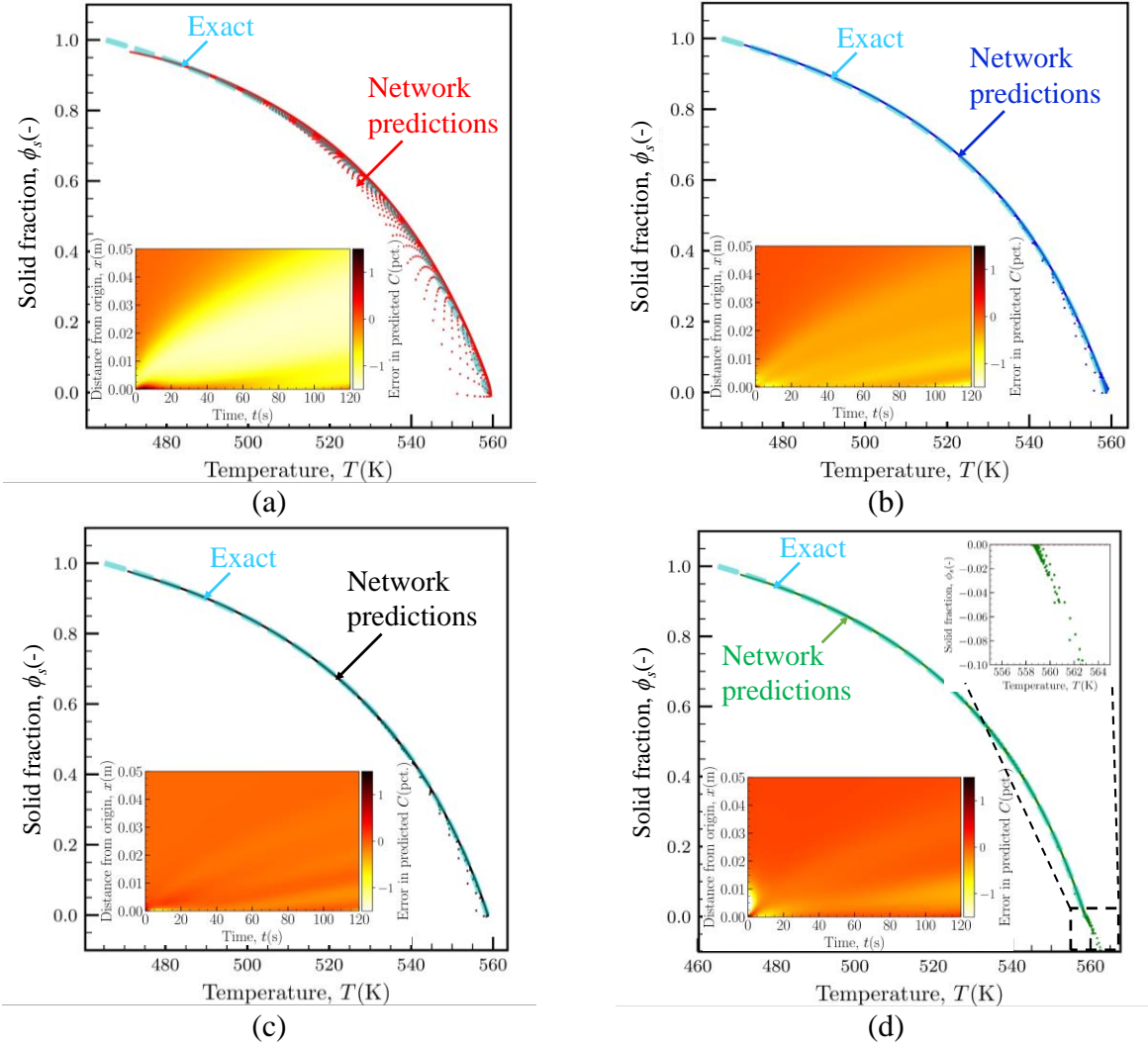

Figure 7. The predicted solid fractions plotted as a function of the predicted temperatures. Plots (a) to (d) show the predictions of four different networks, whose training was displayed in Figure 6 and had $\lambda_0 = 2.5\times10^{-6}$ (a), $1.0\times10^{-5}$ (b), $3.0\times10^{-5}$ (c), and $1.0\times10^{-4}$ (d). The colored insets show the errors in the predicted $C$, defined as $(C - C_0)/C_0$, across the entire spatio-temporal domain. The plot in the top-right inset of (d) is a close up around the liquidus temperature.

### 6.3 A curve-fitting relation for lower/upper limits of the range of optimum learning rates

To present a relation for the range of optimum initial learning rates $\lambda_0$, introduced in Sub-section 6.2, the lower and upper limits of this range need to be first formally defined. To define the lower limit, first, note from Figure 6 that for a $\lambda_0$ below the lower limit, an increase in $\lambda_0$ will decrease the final loss. Therefore, we define the lower limit as the lowest $\lambda_0$ which gives a final loss that is only five times larger than the lowest final loss that can be obtained by using any value for $\lambda_0$. In addition, we define the upper limit as the lowest $\lambda_0$ with which the maximum value of $\Delta loss/loss$ during training exceeds 0.25, where $\Delta loss$ is the difference between the loss function at two consecutive epochs. These definitions are, to some extent, subjective; therefore, the curve-fitting relations that are introduced next should be viewed only as a guideline to search for an appropriate value of $\lambda_0$.

To determine the values of $\lambda_0$ that correspond to the lower and upper limits, one needs to train networks with different values of $\lambda_0$, investigate the final value of the loss and the maximum value of $\Delta loss/loss$ during training, and then identify the two $\lambda_0$ values that correspond to the limits, respectively. A sample of this process is shown in Figure 8. In the figure, the final loss and the maximum of $\Delta loss/loss$ during training of fifteen different networks that all have $D = 10$ and $W = 5$, but were trained using different values of $\lambda_0$ are displayed. The vertical black/blue dotted lines show the lower/upper limits, and they correspond to $\lambda_0 = 4.0 \times 10^{-6}$ and $2.42 \times 10^{-5}$, respectively. Similar results were obtained for networks with other values of $D$ and $W$, but, for the sake of brevity, they are not displayed here. In Figure 9, the $\lambda_0$ values corresponding to the lower and upper limits for networks with different $D$ and $W$ are displayed as a function of $D \times W$ with red and black circles, respectively. The limits determined in Figure 8, which were for networks with $D = 10$ and $W = 5$ and therefore $D \times W = 50$, are enclosed by blue circles. The red and black curves show our fits to the data. Using these fits, the following relation can be suggested for the range of $\lambda_0$ values that should be used

$$0.15 \times 0.075(DW)^{-2.0} < \lambda_0 < 0.075(DW)^{-2.0} \tag{21}$$

The above relation indicates that the lower limit is about seven times smaller than the upper limit, and the range of optimal $\lambda_0$ values shrinks with an increase in $D$ or $W$. The relation also shows that when $D$ or $W$ is increased, $\lambda_0$ should be decreased by an amount that can be expected to be inversely proportional to the square of the increase in $D$ or $W$.

We emphasize that the observation in the present study, the fitting relation proposed in equation (*21*), and the guidelines discussed below the equation are based only the training experiments we performed and lack a fundamental theoretical proof; therefore, they are not guaranteed to remain entirely valid if training is performed using, for example, a different solidification model. Clearly, more studies are needed to clarify this issue. Nonetheless, our main observation, which stated that the predictions of a TTN would be accurate only when the training loss reaches a plateau, and that is achieved only when the learning rate is within an intermediate range of values, can be expected to remain valid even if the training conditions change. In addition, our observations were based on hundreds of training experiments (some not displayed here for the sake of brevity), and the training guidelines we provide can be helpful in future studies by reducing the number of training experiments that need to be performed to find TTNs that give accurate predictions.

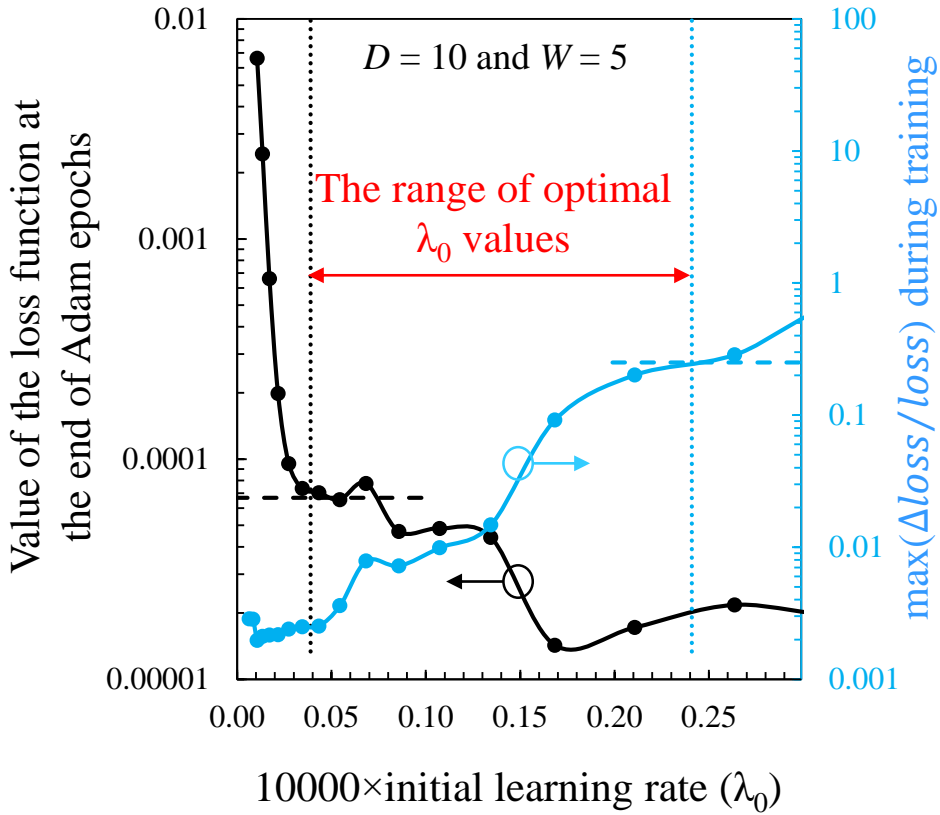

Figure 8. Values of the loss function at the end of Adam epochs (left axis) and the maximum values of the relative changes in loss during training (right axis) for fifteen different networks that were trained using different values of $\lambda_0$.

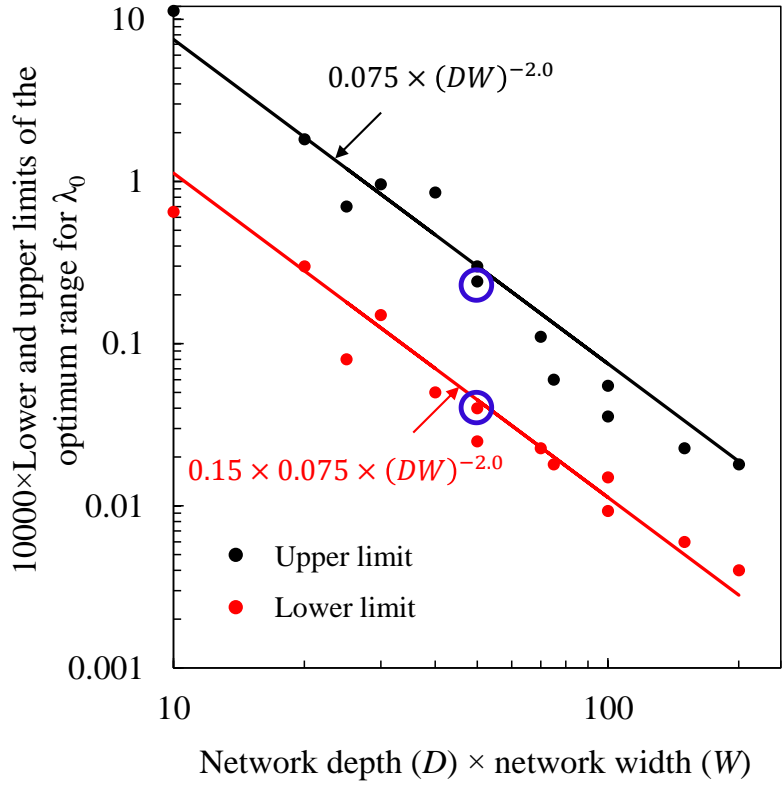

Figure 9. Lower/upper limits (red/black markers) of the range of optimum $\lambda_0$ values plotted as a function of $D \times W$. The solid lines show our curve fits. The limits determined in Figure 8, for networks with $D = 10$ and $W = 5$, are enclosed with blue circles.

# 7 Conclusions

The application of deep neural networks, and more specifically the framework which was founded by Raissi et al. [1]–[3], was extended to the field of alloy solidification modeling. For the first time in the field, TTNs, which simulate a well-known benchmark problem, were introduced. These networks were trained using a theory, which consists of a macroscale model for solidification and the initial/boundary conditions of the benchmark problem. One of the main advantages of TTNs is that they do not need any external data for training or prior knowledge of the solution of the governing equations.

The essential ingredients of the theory-training process were discussed in detail. TTNs with different weights and depths were theory-trained using different learning rates. Their training curves and predictions were analyzed. It was found that only networks that reach a plateau in their training curves are able to provide predictions that agree well with the exact solution of the governing equations and respect the physical limits of the model variables. Reaching that plateau requires the value of the learning rate $\lambda_0$ to be within a specific range. A curve fitting relation for determining the lower/upper limits of that range was provided.

It should be emphasized that this paper presented only the first step in applying deep neural networks in the field of alloy solidification modeling. Our observations and recommendations on how to choose the learning rate were based only on the training experiments we performed in the present study; therefore, they are not guaranteed to hold if, for example, training is performed using a different solidification model. This is mainly because these observations currently lack a rigorous mathematical proof. However, linking the training curve of a TTN to its predictions, as was done in the present study, provides some insight on how to train networks that provide accurate predictions.

There are numerous possibilities to extend the present study in different directions. For example, it can be extended to study three and four-dimensional solidification. Training higher-dimensional datasets will be computationally more expensive, and it would be then interesting to compare the training cost with the cost of performing simulations using a finite volume method. It is also interesting to theory-train using a solidification model that incorporates melt convection.

Acknowledgment

This study was funded by the Deutsche Forschungsgemeinschaft (DFG, German Research Foundation) under Germany's Excellence Strategy – EXC-2023 Internet of Production – 390621612. The support from Ralph Altenfeld of ACCESS e.V. in training the networks on our in-house cluster is also appreciated.

# References

[1]   M. Raissi, P. Perdikaris, and G. E. Karniadakis, "Physics Informed Deep Learning (Part I): Data-driven Solutions of Nonlinear Partial Differential Equations," *arXiv Prepr. arXiv1711.10561*, 2017.

[2]   M. Raissi, P. Perdikaris, and G. E. Karniadakis, "Physics Informed Deep Learning (Part II): Data-driven Discovery of Nonlinear Partial Differential Equations," *arXiv Prepr. arXiv1711.10566*, 2017.

[3]   M. Raissi and G. E. Karniadakis, "Hidden physics models : Machine learning of nonlinear partial differential equations," *J. Comput. Phys.*, vol. 357, pp. 125–141, 2018.

[4]   I. Goodfellow, Y. Bengio, and A. Courville, *Deep Learning*. MIT press., 2016.

[5]   K. Hornik, M. Stinchcombe, and H. White, "Multilayer feedforward networks are universal approximators," *Neural Networks*, vol. 2, no. 5, pp. 359–366, Jan. 1989.

[6]   K. Hornik, "Approximation capabilities of multilayer feedforward networks," *Neural Networks*, vol. 4, no. 2, pp. 251–257, 1991.

[7]   Z. Lu, H. Pu, F. Wang, Z. Hu, and L. Wang, "The expressive power of neural networks: A view from the width," in *Advances in Neural Information Processing Systems*, 2017, vol. 2017–Decem, no. Nips, pp. 6232–6240.

[8]   M. Telgarsky, "Benefits of depth in neural networks," in *Journal of Machine Learning Research*, 2016, vol. 49, no. June, pp. 1517–1539.

[9]   N. Cohen, O. Sharir, and A. Shashua, "On the Expressive Power of Deep Learning: A Tensor Analysis," *J. Mach. Learn. Res.*, vol. 49, no. June, pp. 698–728, Sep. 2015.

[10]  R. Eldan and O. Shamir, "The power of depth for feedforward neural networks," in *Journal of Machine Learning Research*, 2016, vol. 49, no. June, pp. 907–940.

[11]  J. Schmidt, M. R. G. Marques, S. Botti, and M. A. L. Marques, "Recent advances and
38